%% file: Main.tex
\def\IEEEsubmission{0}
\if\IEEEsubmission1
\documentclass[journal,12pt,onecolumn,draftclsnofoot]{IEEEtran}
\else
\documentclass[journal]{IEEEtran}
\fi
\usepackage[utf8]{inputenc}
\usepackage{multicol}
\usepackage{mathtools}
\usepackage{amsthm}

\usepackage{amsfonts}
\usepackage[bookmarksopen=true]{hyperref}
\usepackage{stfloats}
\usepackage{acro}
\usepackage[noadjust]{cite}
\usepackage{multirow}
\usepackage{bm}
\usepackage{enumitem}
\usepackage{amsmath,amssymb}
\usepackage{graphicx}
\usepackage{epstopdf}
\usepackage[table,xcdraw]{xcolor}
\usepackage[geometry]{ifsym}
\usepackage{array}
\usepackage[utf8]{inputenc}
\usepackage[T1]{fontenc}
\usepackage{pifont}

\usepackage{algorithmic}
\usepackage{algorithm}
\usepackage[caption=false,font=normalsize,labelfont=sf,textfont=sf]{subfig}
\usepackage{textcomp}
\usepackage{url}
\usepackage{verbatim}
\usepackage{xcolor}
\usepackage[normalem]{ulem}
\providecolor{added}{rgb}{0,0,1}
\providecolor{deleted}{rgb}{1,0,0}

\def\BibTeX{{\rm B\kern-.05em{\sc i\kern-.025em b}\kern-.08em
		T\kern-.1667em\lower.7ex\hbox{E}\kern-.125emX}}

\input{Acronyms}
\usepackage{cite}

\begin{document}
\title{Scalable Multi-BD Access for OFDM-Based Symbiotic Backscatter Communications}

\author{Fikiri Salum Uledi, Muhammad Bilal Janjua,~\IEEEmembership{Member,~IEEE}, Muhammad Sohaib J. Solaija,~\IEEEmembership{Member,~IEEE}, Cagri Ozgenc Etemoglu, and H\"{u}seyin Arslan,~\IEEEmembership{Fellow,~IEEE}
\thanks{F. S Uledi and H. Arslan are with the Department of Electrical and Electronics Engineering, Istanbul Medipol University, Istanbul, 34810, T\"{u}rkiye (email: fikiri.uledi@std.medipol.edu.tr, huseyinarslan@medipol.edu.tr). 
\newline M. B. Janjua is the R\&D Department, Oredata, Esenler 34220, Istanbul, T\"{u}rkiye, and also with Turk Telekom R\&D Department, Istanbul 34660, T\"{u}rkiye (email: bilal.janjua@ieee.org).
\newline  M.S.J. Solaija is with the Institute of Defence Technologies, Gebze Technical University, Gebze 41400, Kocaeli, T\"{u}rkiye (email: solaija@gtu.edu.tr).
\newline C. O. Etemoglu is with the Turk Telekom R\&D Department, Istanbul 34660, Turkey (email: cagriozgenc.etemoglu@turktelekom.com.tr).}
}

\maketitle
\begin{abstract}
We propose an interference-free multi-backscatter device (BD) access scheme for orthogonal frequency division multiplexing (OFDM)-based symbiotic backscatter communication (SBC). Specifically, we introduce an orthogonal frequency-code spread (OFC) scheme in which the transmitter reserves empty subcarriers for BDs to incorporate their information through frequency shifting, which effectively suppresses direct-link interference (DLI). Since enabling simultaneous access over shared subcarriers introduces inter-backscatter device interference (IBDI), a blockwise orthogonal code-spread is applied to assign each BD to a unique codeword, thereby mitigating IBDI with orthogonal transmissions of all BD signals in the code domain. We develop a non-coherent energy detector whose performance is analyzed in terms of average probability of missed detection (PMD) and sum-rate under Rayleigh fading. The simulation results validate the OFC scheme and demonstrate  low PMD values and high spectral efficiency compared to conventional code-assisted multiple-BD access schemes.
\end{abstract}

\vspace{-1 em}
\section{Introduction}
\label{sec:intro}
\IEEEPARstart{T}{he} massive deployment of ultra-low-power \ac{IoT} devices requires sustainable and energy-efficient connectivity. Consequently, \ac{3GPP} through Release 19 classifies \ac{A-IoT} devices based on energy storage, design complexity, and power consumption. Particularly, fully passive and semi-passive devices (Device~1 and Device~2a) operating around $1~\mu\text{W}$ are categorized as \acp{BD} \cite{3GPP_ref1}. Unlike conventional \ac{IoT} devices, \acp{BD} perform \ac{BC}. A paradigm in which devices modulate and reflect incident \ac{RF} signals from surrounding sources instead of generating their own carrier signal \cite{AmBC_ContemSurv}. Currently, significant efforts are focused on integrating \ac{BC} into existing wireless systems in order to achieve efficient spectrum usage and reduce the infrastructure costs required for external carrier signal sources. In this context, the \ac{SBC} framework has been proposed to enable tighter coordination between \acp{BD} and the carrier source, wherein the primary signal is deliberately designed to accommodate \ac{BD} operations, thereby transforming \acp{BD} from passive listeners into cooperative network elements~\cite{Janjua_SymbioticRad}. {While recent studies further extending symbiotic radio concepts toward intelligent network-level computation and optimization functionalities~\cite{rev2_no_ct}}, the scalability of \ac{SBC} is still severely constrained in multi-\ac{BD} \ac{BC} scenarios due to interference. Specifically, \ac{DLI} which occurs when the strong transmitter (Tx)-receiver (Rx) link dominates and masks the weak \ac{BD} reflected signals, and also \ac{IBDI} arising from concurrent transmissions of multiple \acp{BD}.

These challenges motivate the adoption of waveform designs that inherently provide orthogonality and fine-grained resource control. As \ac{OFDM} is the de facto waveform in modern wireless systems, several studies have exploited its design features to mitigate interference and enable, efficient, interference-aware backscatter communication. Authors in \cite{MultiRider} exploited subcarrier-pattern diversity and midair frequency synthesis to support many concurrent \ac{OFDM} backscatter tags, but it requires dual \ac{RF} chains at the \ac{Rx} and subcarrier-pattern reconstruction plus linear equation solving, which is difficult to integrate into a simple \ac{SR} \ac{Tx} and ultra-low-power tags. Likewise, \cite{CBMA} uses direct sequence spread spectrum (DSSS)-style pseudo-noise (PN) spreading and tag-side power control to handle asynchronous tags and near-far effects, but long PN codes, correlation-based detection, and node-selection feedback increase signaling overhead and tag complexity for accurate modulation. Also the design is tailored to Wi-Fi-backscatter-specific rather than a symbiotic framework that preserves an incumbent cellular \ac{OFDM} downlink unchanged. Similarly, low activity-\ac{CDMA} \ac{BC} scheme in \cite{LA_CDMA_SR} introduces sparsity-aware maximum a posteriori (S-MAP) and sparsity-aware iterative successive interference cancellation (S-SIC) detectors that jointly decode primary and \ac{BD} signals, yet they require accurate cascaded \ac{CSI} and activity statistics which operate on high-dimensional sparse vectors, hence leading to exponential or iterative complexity at a cooperative multi-antenna \ac{Rx}. Furthermore, the random code-assisted \ac{SR} framework in \cite{RandomCode_SR} optimizes primary power and \ac{BD} reflection coefficients using high system \ac{SINR} analysis and block coordinate descent, but assumes knowledge of statistical \ac{CSI} and involves periodic broadcast of optimized coefficients to all \acp{BD}, which is a stringent assumption for dense passive \ac{IoT} deployments. {In,~\cite{rev2_to_ct} direct-link-assisted multiuser \ac{SBC} is studied using joint waveform design and code-aided blind detection for improved spectral efficiency. However, its adoption to \ac{NOMA}, makes the primary and \ac{BD} signals remain coupled at the receiver, which increases receiver complexity.} In \cite{OrthoCodeA}, orthogonal code-spread-assisted \ac{BC} uses time-domain binary sequences for simple correlator-based detection under asynchronous operation, {while~\cite{rev1_lett,rev1_jn} achieve interference-free BC by shifting the reflected signal to non-overlapping spectral regions. However, these schemes are not specifically tailored to \ac{OFDM}-based operation and do not consider joint waveform design, which may limit their practical integration into existing systems.}

To address the challenges of complex multiuser detection, midair frequency synthesis, and extensive \ac{CSI} or control signaling while preserving scalability and hardware efficiency in \ac{OFDM}-based \ac{SBC} systems. In this letter, we revisit our previous work in~\cite{Interference-FreeBC_Janjua} and propose a scalable multi-\ac{BD} access framework particularly an \ac{OFC} modulation scheme. The proposed scheme reuses empty subcarriers with lightweight orthogonal code assignments as \ac{RF} resources for \acp{BD} signals to enable concurrent multi-\ac{BD} access. The main contributions are summarized as follows:

\begin{figure}[t]
    \centering
    \includegraphics[width=\linewidth]{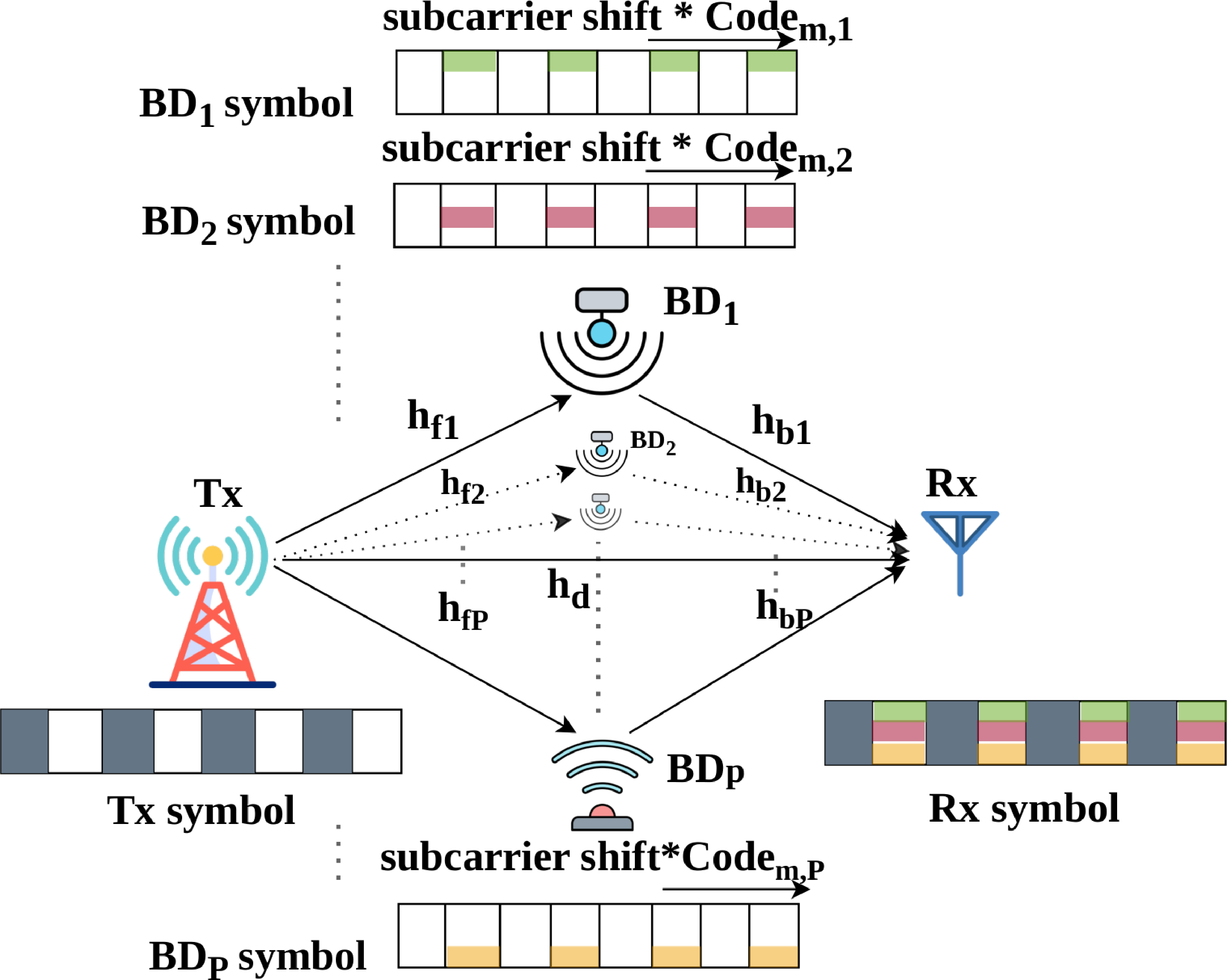}

    \vspace{-0.3cm}
    \caption{The proposed multi-BD OFDM-based \ac{SR} system model.}
    \label{fig:image1}
\end{figure}

\begin{itemize}
  \item We propose an \ac{OFC} multi-\ac{BD} access scheme where \ac{Tx} transmits an \ac{OFDM} signal with an empty subcarrier allocated after each data subcarrier. \acp{BD} modulate the incoming \ac{OFDM} symbol by applying \ac{OFC} to simultaneously suppress both \ac{DLI} and \ac{IBDI}.
  \item We derive a low-complexity non-coherent detector for multi-\ac{BD} data detection and perform analytical analysis for \ac{PMD} and the sum-rate.
  \item We validate the proposed scheme through simulations and provide quantitative analysis, indicating both detection and sum-rate performance gains compared to the prior art.
\end{itemize}

\vspace{-0.35cm}
\section{Signal and System Model}
\label{sec:sysmod}
We consider a cooperative \ac{OFDM}-based \ac{SBC} system in Fig.~\ref{fig:image1}, where \ac{Tx} transmits an \ac{OFDM} signal to \ac{Rx} through the direct link channel, $h_{\mathrm{d}}$ while simultaneously serving the $P$ \acp{BD} classified as \ac{3GPP} Device~1. Each $\mathrm{BD}_p$ modulates the $m^{th}$ symbol of the incident \ac{OFDM} signal and reflects to \ac{Rx} through cascaded forward $h_{\mathrm{f}p}$ and backscatter $h_{\mathrm{b}_p}$ links. All links are modeled as multipath Rayleigh fading. {Through a microcontroller-driven switching network, \(\mathrm{BD}_p\) employs low-complexity impedance modulation leading to single-sideband frequency shifting and orthogonal code-domain modulation. Specifically, by varying the load impedance $Z_{l_p}$ relative to the antenna impedance $Z_{a_p}$, it controls its reflection coefficient $\alpha_p=\frac{Z_{l_p}-Z_{a_p}^{*}}{Z_{l_p}-Z_{a_p}}$ where information is encoded through reflecting ($Z_{l_p}=Z_{a_p}$) and absorbing ($Z_{l_p}\neq Z_{a_p}$) states.}

Initially, \ac{Tx} transmits a block of $M$  \ac{OFDM} symbols with $N$ subcarriers, where in each \ac{OFDM} symbol the empty subcarrier is allocated after the data subcarrier to create spectral resources for the \ac{BD} signals after going through a frequency shift. Therefore, the corresponding \ac{Tx} subcarrier mapping is given as

\vspace{-0.3cm}
\begin{equation}
    X[k] =
\begin{cases} 
s[q], & k = 2q\\
0, & \text{otherwise}
\end{cases}~,
\label{eqn:x[k]}
\end{equation}
where $q \in \mathbb{Z}^+$ with $q \leq N_d-1$, and $s[q] = [s[0], s[1], \ldots, s[N_d-1]]$ denotes the primary data-symbol vector of length $N_d = \frac{N}{2}$ with the $N_d$ being the number of data subcarriers. Applying \ac{IDFT} on~(\ref{eqn:x[k]}), the time-domain \ac{OFDM} signal is obtained as

\vspace{-0.5cm}
\begin{equation}
x[n] = \frac{1}{\sqrt{N}} \sum_{k=0}^{N-1} X[k] e^{j \frac{2\pi k n}{N}}~, \quad 0 \leq n \leq N-1~.
\end{equation}
To enforce circular convolution and \ac{ISI} mitigation, the \ac{CP}-appended waveform, denoted by $x_{\mathrm{cp}}[n]$, is formed by adding the \ac{CP} of length $N_{cp}$.

\vspace{-0.1cm}
\section{Proposed \ac{OFC} Modulation Scheme}
\label{sec:propzd_scheme}
Since \ac{OFDM} distributes data symbols across multiple subcarriers~\cite{farhang2016ofdm}, to achieve \ac{DLI}-and \ac{IBDI}-free multi-\ac{BD} access, this article proposes \ac{OFC}; a hybrid passive modulation scheme that simultaneously performs  frequency shift and orthogonal code spread. Specifically, each $\mathrm{BD}_p$ employs a controllable switching of $Z_{\mathrm{l}_p}$ to impose a subcarrier shift associated with an orthogonal code depending on its information bit, $b_p$ to be transmitted. The subcarrier shift maps the $\mathrm{BD}_p$ signal to the reserved empty subcarriers, which enables orthogonal frequency reception between all the \acp{BD} signals and the direct link signal, thus ensuring \ac{DLI}-free \ac{BC} while orthogonal codes enable orthogonal reception between all the \acp{BD} signals, thereby \ac{IBDI}-free \ac{BC} is achieved. In particular, for the proposed scheme to transmit $b_p$ = `1'; an active state, $\mathrm{BD}_p$ modulates the incoming \ac{OFDM} signal by switching $Z_{\mathrm{l}_p}$ between reflecting and non-reflecting states at a rate that induces a single subcarrier shift multiplied by an orthogonal code, $C_{m,p}$, where $m$ is the symbol index. In contrast, to transmit $b_p$ = `0', an inactive state, $\mathrm{BD}_p$ maintains $Z_{\mathrm{l}_p}$ in a fixed state, also without any operation with $C_{m,p}$. This operation can mathematically be expressed as

\vspace{-0.3cm}
\begin{equation}
\beta_p[n] =
\begin{cases}
0, & b_p=0\\
C_{m,p}\, e^{j 2\pi f_\xi n}, & b_p=1
\end{cases}~,
\end{equation}
where $f_\xi$ = $\xi\Delta{f}$ with $\xi = 1$ denoting subcarrier shifting index while $\Delta{f}$ representing the subcarrier spacing in the \ac{OFDM} signal. Therefore, the corresponding \ac{BD} signal is obtained as $x_{\mathrm{b}_p}[n] = (h_{\mathrm{f}p}[n] * x_{\mathrm{cp}}[n]) \,\alpha_p\, \beta_p[n]$
with $*$ denoting a circular convolution.

\vspace{-0.1cm}
\section{Detection and Performance Analysis}
\label{sec:Dectperf}
At \ac{Rx}, the received signal of the $m^{th}$ block is expressed as the composite of all the $P$ \acp{BD} signals and the direct link signal, expressed as

\vspace{-0.7cm}
\begin{equation}
\label{eqn:rec_sig}
y[n]= x[n] * h_{\mathrm{d}}[n] +\sum_{p=1}^P (h_{\mathrm{b}_p}[n] *x_{\mathrm{b}_p}[n])+ w[n]~,
\end{equation}
where $w[n]$ represents an \ac{AWGN}. After \ac{CP} removal and \ac{DFT} operations, the received signal on the data subcarriers, $\hat{k}_{\mathrm{d}}$, is expressed as $\hat{Y}_{\mathrm{d}}[\hat{k}_{\mathrm{d}},m] = H_{\mathrm{d}}[\hat{k}_{\mathrm{d}},m] X[\hat{k}_{\mathrm{d}},m]$. The empty subcarriers, $\hat{k}_p$ are now mapped with a signal composed of all the \acp{BD} data, which is expressed as $\hat{Y}_{\mathrm{b}}[\hat{k}_p,m] = \sum_{p=1}^P\alpha_p H_{\mathrm{f}_p}[\hat{k}_p,m] X[\hat{k}_p] H_{\mathrm{b}_p}[\hat{k}_p,m]$. Thus primary and all the $P$ \acp{BD} signals occupy disjoint subcarrier sets while multiplied by orthogonal blockwise $C_{m,p}$. {Given that  $H_{\mathrm{d}}$ and $H_{\mathrm{f}_p}$ are modeled as multi-tap Rayleigh fading channels in time domain, they result in frequency-selective per-subcarrier coefficients after OFDM demodulation. Following the short $\mathrm{BD}_p$-to-Rx distance, $H_{\mathrm{b}_p}$ is then modeled as a single-tap channel, as commonly adopted in low-power \ac{BC} analysis~\cite{Interference-FreeBC_Janjua, elmossallamyCHANNEL}. Hence, on each null subcarrier, the Rx observes an approximately narrowband effective cascaded channel, and the detector analysis is performed using the corresponding aggregated per-subcarrier channel coefficient. For analytical tractability, the $H_{\mathrm{f}_p}$ power gain is approximated by its average value, i.e., $G_f \approx |H_{fp}[\hat{k}_p,m]|^2 \approx \mathbb{E}\{|H_{fp}[\hat{k}_p,m]|^2\}$, while the remaining randomness is captured by $z \triangleq |H_{bp}[\hat{k}_p,m]|^2$. This avoids intractability arising from the product of multiple random channel components in the cascaded link, thereby enabling a tractable characterization of the detection statistic.}

To recover the $\mathrm{BD}_p$ signal, \ac{Rx} cancels the multiplicative effect of the data symbols to remove the associated randomness, $Y_{\mathrm{b}}[\hat{k}_p,m] = \frac{\tilde{Y}_{\mathrm{b}}[\hat{k}_p,m]}{X[\hat{k}_{\mathrm{d}},m]}$, it then performs code correlation to disperse across the $M$ blocks and non-coherent energy detection to get the \ac{BD} information. The corresponding \ac{BD} signals at $\hat{k}_p$ on the $m^{th}$ block can be written as

\vspace{-0.35cm}
\begin{equation}
\label{eqn:BDsigfrd}
    Y_{\mathrm{b}}[\hat{k}_p,m] =
    \begin{cases}
        W[\hat{k}_p,m], & b_p=0\\
        \tilde{Y}_{\mathrm{b}}[\hat{k}_p,m] + W[\hat{k}_p,m], & b_p=1
    \end{cases}~,
\end{equation}
where $W[\hat{k}_p,m]\sim\mathcal{CN}(0,\sigma_W^2)$ is \ac{AWGN}. 
To exploit the code orthogonality across $M$ blocks in each $\hat{k}_p$ from the $N_b$ number of empty subcarriers, the \ac{Rx} performs code correlation for the entire set of $\hat{k}_p$, expressed as

\vspace{-0.3cm}
\begin{equation}
\label{eqn:Zm}
    \Phi_p =\sum_{\hat{k}_p \in \mathcal{K}_{b}}\sum_{m=1}^{M} C_{m,p}\, Y_{\mathrm{b}}[\hat{k}_p,m]~,
\end{equation}
where $\mathcal{K}_{b}$ is a set of empty subcarriers. 

The $\Phi_p$ in \eqref{eqn:Zm} is the code-correlated combining output which acts as a sufficient statistic for isolating the contribution of $\mathrm{BD}_p$ across the $M$ blocks and the $\mathcal{K}_{b}$. Based on $\Phi_p$ we construct the non-coherent energy metric $E_p \triangleq |\Phi_p|^2$ which serves as a test statistic for the detection of activity \ac{BD}. Specifically, the decision problem is formulated as a binary hypothesis test on the presence of the backscatter component from $\mathrm{BD}_p$, i.e., $\mathcal{H}_0$ (inactive state, $b_p=0$) versus $\mathcal{H}_1$ (active state, $b_p=1$). In the sequel, we characterize the distribution of $E_p$ under both hypotheses and compute the detection threshold accordingly.

\begin{itemize}
\item \textbf{Inactive \ac{BD} state, $\mathcal{H}_0$}:\\ It is the null hypothesis, considered when $\Phi_p$ contains only the aggregated noise terms over the $\mathcal{K}_{b}$ in $M$ blocks, i.e.,

\vspace{-0.35cm}
\begin{equation}
    \Phi_p[\hat{k}_p]\triangleq \sum_{m=1}^{M} C_{m,p}\,W[\hat{k}_p,m]\sim \mathcal{CN}\big(0,\sigma_W^2\big)~.
\end{equation}

The corresponding normalized energy statistics in this hypothesis, $\frac{2}{M\sigma_W^2} E_p \,\big|\, \mathcal{H}_0 \sim \chi^2_{2N_bM}$ is later expressed as
\begin{equation}
U_p \triangleq \frac{2}{M\sigma_W^2}\,E_p \,\Big|\,\mathcal{H}_0 \sim \chi^2_{2N_bM}~.
\label{eq:Ep_H0}
\end{equation}

\item \textbf{Active \ac{BD} state, $\mathcal{H}_1$}:\\ It is the alternative hypothesis considered when $\Phi_p$ on each $\hat{k}_p$ is obtained after code-combining over $M$ blocks, yielding a complex Gaussian variable with a non-zero mean determined by the cascaded link $H_{\mathrm{f}_p}[\hat{k}_d,m]H_{\mathrm{b}_p}[\hat{k}_p,m]$. The detector then forms the energy metric by summing across the $N_b$ reserved subcarriers. Conditioning on $G_f$ and $z$, the normalized energy statistic, $U_p \triangleq \frac{2}{M\sigma_W^2}\,E_p$ follows a non-central chi-square distribution

\vspace{-0.5cm}
\begin{equation}
U_p \,\big|\, z \sim \chi'^2_{2N_bM}\!\big(\lambda_p(z)\big)~,
\qquad
\lambda_p(z)=2N_b\,\gamma_p(z)~,
\label{eq:lambda_H1}
\end{equation}
where the effective post-combining \ac{SNR} is defined as

\vspace{-0.3cm}
\begin{equation}
\gamma_p(z) \triangleq \frac{E_s\,\alpha_p^{2}\,G_f\, z}{\sigma_W^{2}}~,
\label{eq:gamma_H1}
\end{equation}
where $E_s=\mathbb{E}\{|X[\hat{k}]|^2\}$ denoting the average per-subcarrier symbol energy of the primary OFDM signal.
\end{itemize}

The detection performance of the proposed scheme is obtained by evaluating \ac{PMD} for a threshold $\eta$ which is determined based on the given \ac{PFA}. Under equal prior probabilities for $\mathcal{H}_0$ and $\mathcal{H}_1$, the \ac{PMD} is expressed as $P_{\mathrm{MD}}(\eta; z) = \Pr(E_P \le \eta \mid \mathcal{H}_1, z)$ while the \ac{PFA} expressed as $P_{\mathrm{FA}}(\eta) = \Pr(E_p > \eta \mid \mathcal{H}_0)$. Given the distribution in \eqref{eq:Ep_H0}, \ac{PFA} is further expressed as

\vspace{-0.3cm}
\begin{equation}
    P_{\mathrm{FA}}(\eta) = 1 - F_{\chi^2_{2N_bM}}\!\left(\frac{\eta}{M\sigma_W^2}\right)~,
    \label{eq:ga1}
\end{equation}
where $F_{\chi^2_{2N_bM}}(\cdot)$ is the \ac{CDF} of a chi-square random variable with $2N_bM$ degrees of freedom. {This explicitly indicates that the threshold depends on the noise variance, the number of reserved subcarriers, and the code-combining length, thereby clarifying the design tradeoff between reliability and resource overhead}. For a target $P_{\mathrm{FA}} =\epsilon$, $\eta$ is then chosen as $\eta = \frac{M\sigma_W^2}{2}\,\chi^2_{2N_b,\,1-\epsilon}$, where $\chi^2_{2N_bM,\,1-\epsilon}$ is the $(1-\epsilon)$-quantile of $\chi^2_{2N_bM}$. Conditioned on $z$, the $P_{\mathrm{MD}}$ is then obtained as

\vspace{-0.3cm}
\begin{equation}
    P_{\mathrm{MD}}(\eta; z) 
    = F_{\chi'^2_{2N_bM}(\lambda_p(z))}\!\left(\frac{\eta}{M\sigma_W^2}\right)~,
    \label{eq:ga1ast}
\end{equation}
where $F_{\chi'^2_{2N_bM}(\lambda_p)}(\cdot)$ is the \ac{CDF} of a non-central chi-square distribution with $2N_bM$ degrees of freedom and non-centrality parameter $\lambda_p$. {To this end, substituting components of~\eqref{eq:gamma_H1} and $\eta$ into~\eqref{eq:ga1ast} directly reveals the role of effective post-combining \ac{SNR}, code-domain accumulation, and fading realization in determining \ac{PMD}. The detector is therefore characterized conditionally and the performance is obtained by averaging over the random cascaded backscatter gain.}

Since $z=|H_{\mathrm{b}_p}[\hat{k}_p,m]|^2$ is exponentially distributed with $f_Z(z)=e^{-z}$ for $z\ge 0$, therefore, the estimated $P_{\mathrm{MD}}$ can be expressed as

\vspace{-0.5cm}
\begin{equation}
\label{eq:avg_kmd}
    \bar{P}_{\mathrm{MD}}(\eta)
    = \int_{0}^{\infty} 
    F_{\chi'^2_{2N_bM}(\lambda_p(z))}\!\left(\frac{\eta}{M\sigma_W^2}\right)
    e^{-z}\,dz~.
\end{equation}
{In general, \eqref{eq:avg_kmd} does not admit a simple closed-form expression; therefore, it is computed via standard numerical integration using the non-central chi-square \ac{CDF}. This form is retained to preserve generality and explicit parameter dependence, consistent with related \ac{OFDM}-based \ac{BC} detection analyses~\cite{elmossallamyINTERGR1,elmossallam_frshiftOPJ}.}
\begin{figure*}[t]
\centering

\begin{minipage}[t]{0.25\textwidth}
    \centering
    \includegraphics[width=\linewidth]{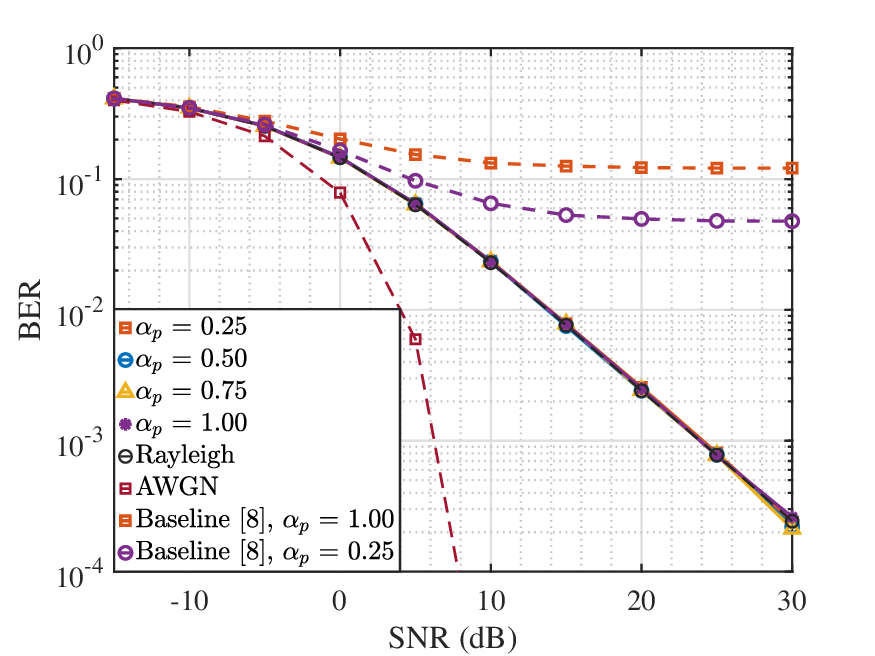}

    \vspace{-0.35cm}
    \caption{BER of primary signal.}
    \label{fig:BER}
\end{minipage}\hfill
\begin{minipage}[t]{0.25\textwidth}
    \centering
    \includegraphics[width=\linewidth]{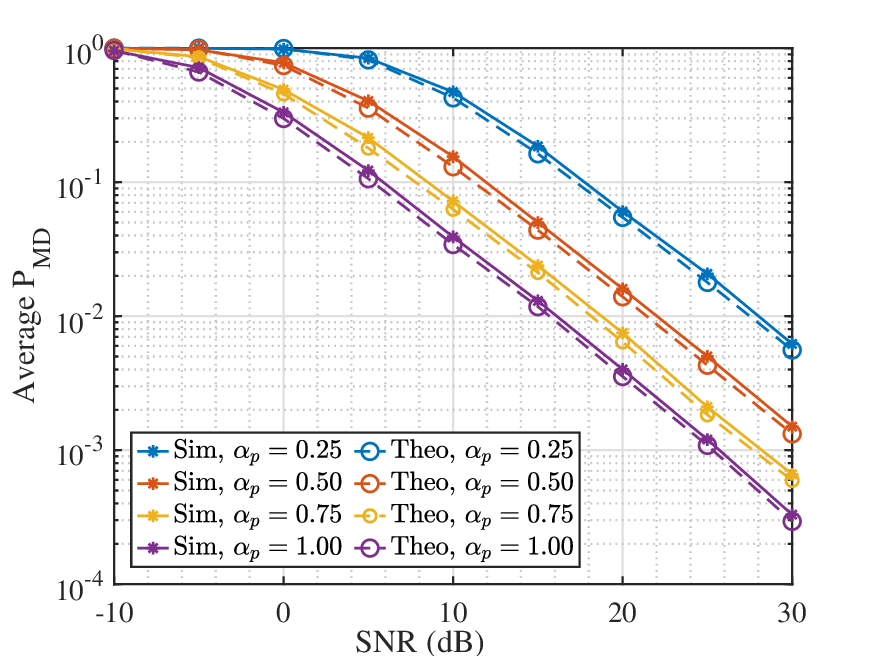}
    
    \vspace{-0.35cm}
    \caption{BD detectoin performance.}
    \label{fig:PMD_A}
\end{minipage}\hfill
\begin{minipage}[t]{0.25\textwidth}
    \centering
    \includegraphics[width=\linewidth]{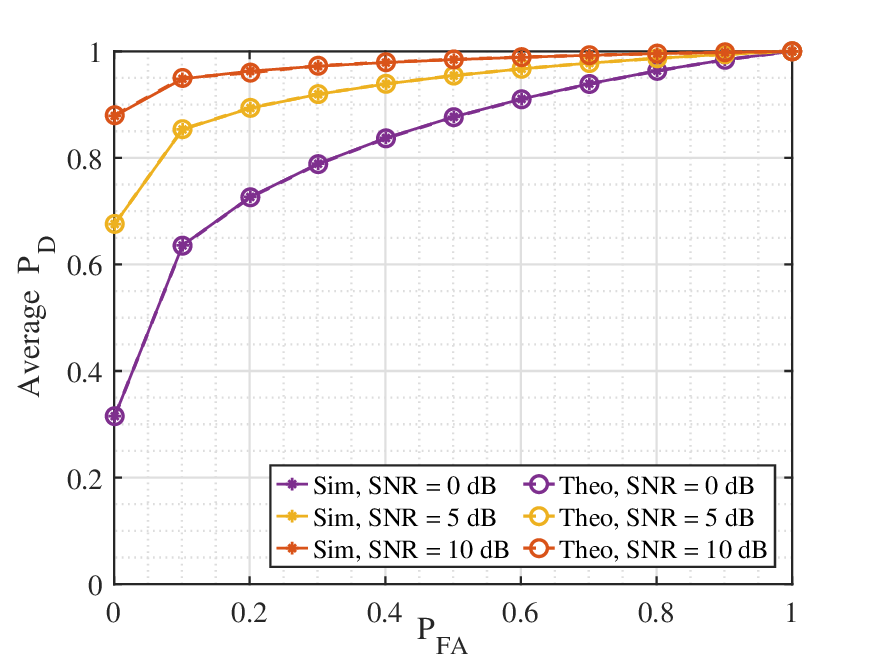}

    \vspace{-0.35cm}
    \caption{BD signal detection ROC.}
    \label{fig:ROC}
\end{minipage}\hfill
\begin{minipage}[t]{0.25\textwidth}
    \centering

    \includegraphics[width=\linewidth]{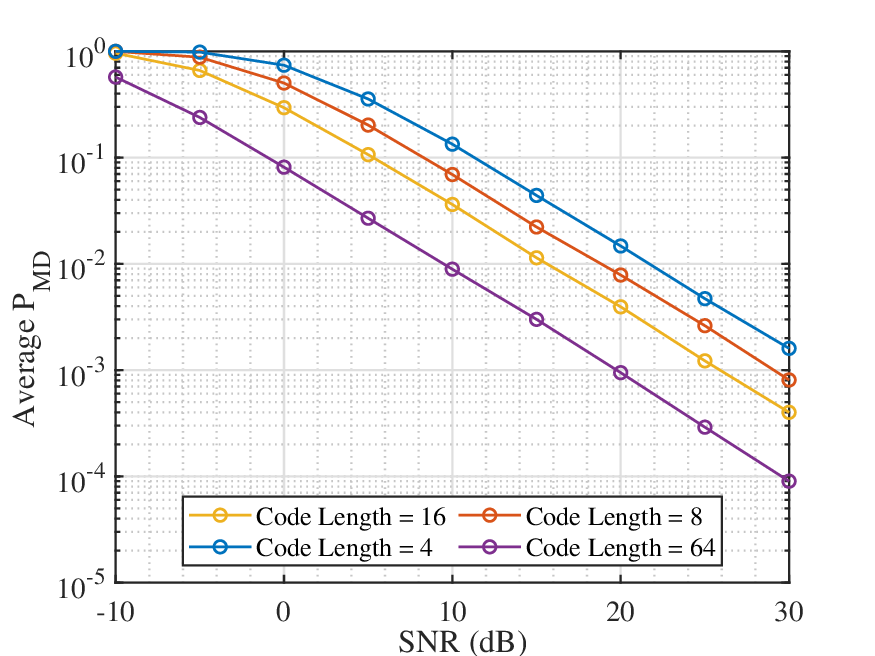}

    \vspace{-0.35cm}
    \caption{Impact of code on detection.}
    \label{fig:CDL}
    
\end{minipage}

\end{figure*}

To quantify spectral utilization in the proposed scheme, we evaluate the aggregate throughput as the sum of the primary-link and BD-link rates accounting for factors, including $\alpha_p$ and the residual interference terms (\ac{IBDI} and \ac{DLI}). Since $\mathrm{BD}_p$ signals are confined to empty subcarriers, at \ac{Rx}, the primary data is free from \ac{BD}-induced interference, hence, the primary-link throughput is \ac{SNR} limited, mathematically expressed as

\vspace{-0.5cm}
\begin{equation}
\mathcal{S}_{Rx}=
\sum_{\hat{k}_d\in\mathcal{K}_d}\Delta f\,
\log_2\!\left(1+\frac{E_s|H_d[\hat{k}_d,m]|^2}{\sigma_W^2}\right)~,
\label{eq:SRx_fix}
\end{equation}
where $\mathcal{K}_{d}$ is a set of primary data subcarriers. Since all BDs share the same set of reserved empty subcarriers and are separated via blockwise orthogonal codes, the aggregate BD throughput can be written in terms of the effective post-despreading SINR. Specifically,

\vspace{-0.3cm}
\begin{equation}
\mathcal{S}_{bd}
= R_c \sum_{p=1}^{P}\sum_{k \in \mathcal{K}_{b}}
\Delta f \, \log_2\!\bigl(1 + \gamma_p(z) )~,
\label{eq:R_bd}
\end{equation}
where $R_c=1/M$ captures code-rate overhead.~{Equation \eqref{eq:R_bd} highlights that, under the proposed \ac{OFC} structure, the \ac{BD}-link throughput is determined by the effective post-despreading \ac{SINR} after \ac{DLI} suppression in the frequency domain and \ac{IBDI} mitigation in the code domain}. The total sum-rate then follows as $\mathcal{S}_{T}=\mathcal{S}_{Rx}+\mathcal{S}_{bd}$ which captures the end-to-end throughput of the proposed system, reflecting how the design parameters combine as propagation effects to determine net data delivery across all links  under prevailing channel conditions, noise levels, and interference.

\begin{table}[!h]
  \centering

  \vspace{-0.4cm}
  \caption{Key simulation parameters.}
  \label{tab:sim_paramt_freco_multibd}
  \footnotesize
  \setlength{\tabcolsep}{16pt}
  \renewcommand{\arraystretch}{0.9}
  \setlength{\extrarowheight}{0.9pt}
  \begin{tabular}{|l|c|}
    \hline
    \textbf{Parameter} & \textbf{Value} \\
    \hline
    DFT size, $N$ & $64$ \\
    \hline
    Cyclic prefix length, $N_{\mathrm{cp}}$ & $8$ \\
    \hline
    Subcarrier spacing, $\Delta f$ & $15~\mathrm{kHz}$ \\
    \hline
    Number of BDs, $P$ & $3$ \\
    \hline
    Code length & $16$ \\
    \hline
    Reflection coefficient, $\alpha_p$ & $1$ \\
    \hline
    Target \ac{PFA}, $\epsilon$ & $10^{-3}$ \\
    \hline
    SNR range & $-15$ to $30$~dB \\
    \hline
  \end{tabular}
\end{table}

\vspace{-0.5cm}
\section{Simulation Results}
\label{sec:Sim_results}
This section provides a numerical evaluation of the proposed schemes. Unless otherwise stated, the key simulation parameters are summarized in Table~\ref{tab:sim_paramt_freco_multibd}. The proposed approach is benchmarked against the baseline in~\cite{RandomCode_SR}, with an emphasis on detection and sum rate performance.

\figurename~\ref{fig:BER} presents the \ac{BER} of the primary link versus \ac{SNR} for different $\alpha_p$ values. The results show a monotonic decrease in \ac{BER} as \ac{SNR} increases. For all $\alpha_p$, the curves remain superimposed, confirming that the mapping of \acp{BD} signals strictly to empty subcarriers effectively preserves the performance of the primary link while suppressing \ac{DLI} to \acp{BD} transmissions. The obtained \ac{BER} performance is further compared with the baseline in~\cite{RandomCode_SR} for $\alpha_p \in \{0.25, 1\}$. The results show that the proposed \ac{OFC} scheme consistently outperforms the baseline, which exhibits a floor \ac{BER} at moderate to high \ac{SNR} due to residual interference from \acp{BD} signals, where in higher $\alpha_p$ more interference is experienced than in lower ones caused by high \acp{BD} signals power.

\figurename~\ref{fig:PMD_A} presents the average $P_{\mathrm{MD}}$ of the proposed non-coherent detector versus \ac{SNR} for different $\alpha_p$ values. The results indicate that the average $P_{\mathrm{MD}}$ decreases as \ac{SNR} increases for all $\alpha_p$. Moreover, higher $\alpha_p$ values significantly provide lower average $P_{\mathrm{MD}}$ values, because higher $\alpha_p$ leads to a higher power of the backscatter signal than lower  $\alpha_p$ values. For instance, at $30$ dB, increasing $\alpha_p$ from $0.25$ to $1$ substantially reduces the average $P_{\mathrm{MD}}$ from nearly $10^{-2}$ to approximately $10^{-4}$, which justifies that strong $\mathrm{BD}_p$ reflections significantly enhance detection of the \ac{BD} signals.

\figurename~\ref{fig:ROC} presents the \ac{ROC} of the proposed non-coherent detector for $\alpha_p=0.25$ for \ac{SNR} values $\in \{0, 5, 10\}$~dB . For a given $P_{\mathrm{FA}}$, the average detection probability $P_{\mathrm{D}}$ increases monotonically with \ac{SNR}, reflecting the improved detectability as the noise power decreases. For example, at $P_{\mathrm{FA}}=0.2$, increasing the \ac{SNR} from 0~dB to 10~dB increases the average $P_{\mathrm{D}}$ from $0.7$ to $0.98$, while at $P_{\mathrm{FA}}=0.4$ it improves from $0.84$ to $0.99$. Moreover, the simulated curves match closely with the corresponding theoretical values, thus validating the analytical model and confirming the $P_{\mathrm{FA}}$-$P_{\mathrm{D}}$ trade-off.

\begin{figure}[t]
  \centering
  \begin{minipage}[t]{0.4\textwidth}
    \centering
    \includegraphics[width=\linewidth]{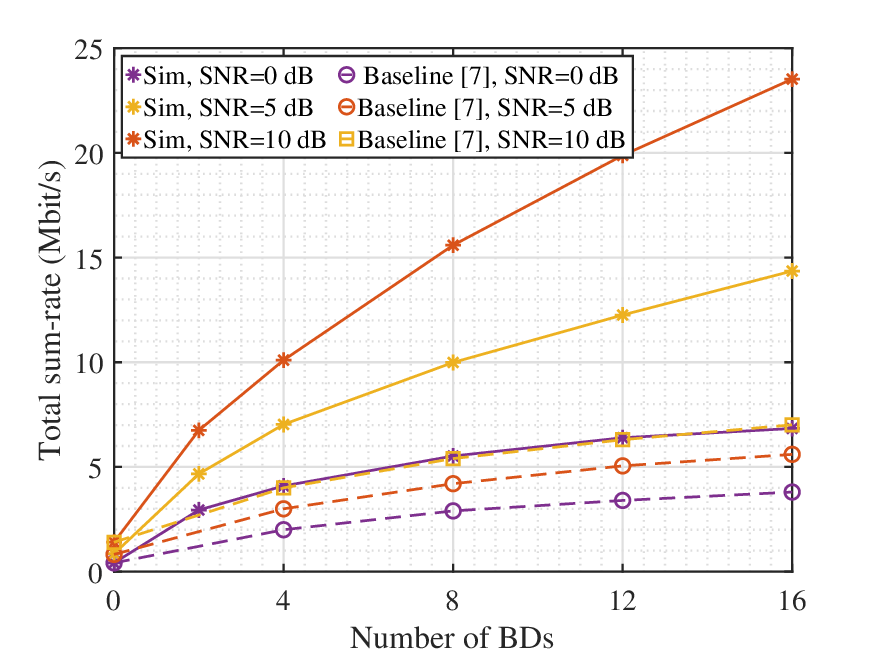}

    \vspace{-0.4cm}
    \caption{Sum rate for different number of BDs.}
    \label{fig:SE_SNRBDs}
  \end{minipage}\hfill
\end{figure}

\figurename~\ref{fig:SE_SNRBDs} provides the total sum-rate, $\mathcal{S}_{T}$ versus the number of \acp{BD} for different \ac{SNR} values. The results show that the $\mathcal{S}_{T}$ increases  as more \acp{BD} are activated, with the larger gains becoming more pronounced at higher \acp{SNR}, since each additional \ac{BD} contributes significantly to the aggregate backscattered throughput and therefore to the total throughput. Compared to the baseline in~\cite{RandomCode_SR}, the proposed design consistently achieves a higher sum-rate and more evident scaling as the number of \acp{BD} grows, especially at moderate-to-high \ac{SNR}. This is because the proposed scheme mitigates both \ac{DLI} and \ac{IBDI}, thereby improving the effective \ac{SINR} and increasing the sum rate.

\figurename~\ref{fig:CDL} presents the average $P_{\mathrm{MD}}$ of the proposed non-coherent detector versus \ac{SNR} for different code lengths. For all code lengths, the average $P_{\mathrm{MD}}$ decreases consistently with increasing \ac{SNR}. Moreover, for a given \ac{SNR}, longer codes generally achieve a lower average $P_{\mathrm{MD}}$ than the shorter ones, because longer codes result in the higher processing gain enabled by correlation-based combining diversity over multiple \ac{OFDM} symbols than shorter codes. For example, in the high-\ac{SNR} regime, e.g., $20$-$30$~dB, code length = 4 achieves high $P_{\mathrm{MD}}$ values, on the order of $10^{-2}$ to $10^{-3}$, while at code length = 64 the values of $P_{\mathrm{MD}}$ are on the order of $10^{-3}$ to $10^{-4}$, indicating a significant improvement.

\vspace{-0.1cm}
\section{Conclusion}
\label{sec:conclusion}
This article proposes an \ac{OFC} multi-\ac{BD} access scheme for interference-free \ac{BC} in multi-\ac{BD} \ac{OFDM}-based \ac{SBC} systems. By jointly leveraging frequency shifting and blockwise orthogonal code spread over dedicated empty \ac{OFDM} subcarriers, the proposed scheme achieves \ac{DLI}-free and \ac{IBDI}-free multi-\ac{BD} access. A non-coherent detector is developed and analytically analyzed. Building on this analysis, the results indicate that increasing the code length or the reflection strength of the BDs significantly enhances both the \ac{PMD} and sum-rate performance. However, the \ac{BER} performance of the primary link remains unaffected. Moreover, the proposed scheme achieves favorable spectral efficiency, making it well suited for massive \ac{IoT} scenarios. Future work includes extending the current analytical detector to a learning-based detector and validating the scheme on a hardware testbed under practical impairments.

\vspace{-0.1cm}
\bibliography{IEEEfull,references}
\bibliographystyle{IEEEtran}
\end{document}

%% file: Acronyms.tex
\DeclareAcronym{OOK}{
short = OOK,
long = on-off keying
}

\DeclareAcronym{OFSK}{
  short = OFSK,
  long = orthogonal frequency shift keying
}

\DeclareAcronym{OFDM}{
short = OFDM,
long = orthogonal frequency division multiplexing
}

\DeclareAcronym{BC}{
short = BC,
long = backscatter communication
}

\DeclareAcronym{CSI}{
short = CSI,
long = channel state information
}

\DeclareAcronym{BS}{
  short = BS,
  long = base station
}

\DeclareAcronym{RFID}{
short = RFID,
long = radio frequency identification 
}

\DeclareAcronym{BD}{
short = BD,
long = backscatter device 
}

\DeclareAcronym{SR}{
short = SR,
long = symbiotic radio 
}

\DeclareAcronym{Tx}{
short = Tx,
long = source transmitter
}

\DeclareAcronym{SBC}{
short = SBC,
long = symbiotic backscatter communication
}

\DeclareAcronym{CP}{
short = CP,
long = cyclic prefix
}

\DeclareAcronym{FSK}{
short = FSK,
long = frequency shift keying
}

\DeclareAcronym{DLI}{
short = DLI,
long = direct-link interference
}

\DeclareAcronym{Rx}{
short = Rx,
long = receiver
}

\DeclareAcronym{SIC}{
short = SIC,
long = successive interference cancellation
}

\DeclareAcronym{ISI}{
short = ISI,
long = inter-symbol interference
}

\DeclareAcronym{MCU}{
short = MCU,
long = microcontroller unit
}

\DeclareAcronym{CFO}{
short = CFO,
long  = carrier frequency offset
}

\DeclareAcronym{ICI}{
short = ICI,
long  = inter-carrier interference
}

\DeclareAcronym{3GPP}{
short = 3GPP,
long  = 3rd generation partnership project
}

\DeclareAcronym{IoT}{
short = IoT,
long  = internet of things
}

\DeclareAcronym{A-IoT}{
short = A-IoT,
long  = ambient IoT
}

\DeclareAcronym{5G-NR}{
short = 5G-NR,
long  = 5G new radio
}

\DeclareAcronym{PMD}{
short = PMD,
long  = probability of missed detection
}

\DeclareAcronym{PD}{
short = PD,
long  = probability of detection
}

\DeclareAcronym{PFA}{
short = PFA,
long  = probability of false alarm
}

\DeclareAcronym{SNR}{
short = SNR,
long  = signal-to-noise ratio
}

\DeclareAcronym{ROC}{
short = ROC,
long  = receiver operating characteristic
}

\DeclareAcronym{BER}{
short = BER,
long  = bit error rate
}

\DeclareAcronym{SE}{
short = SE,
long  = spectral efficiency
}

\DeclareAcronym{SINR}{
short = SINR,
long  = signal-to-noise interference ratio
}

\DeclareAcronym{AWGN}{
short = AWGN,
long  = additive white Gaussian noise
}

\DeclareAcronym{RF}{
short = RF,
long  = radio-frequency
}

\DeclareAcronym{Nb}{
short = Nb,
long  = number of null subcarriers dedicated for SR-BC
}
\DeclareAcronym{Nd}{
short = Nd,
long  = number of subcarriers for primary data transmission
}

\DeclareAcronym{Ncp}{
short = Ncp,
long  = CP length
}

\DeclareAcronym{DFT}{
short = DFT,
long  = discrete Fourier transform
}

\DeclareAcronym{PDF}{
short = PDF,
long  = probability density function
}

\DeclareAcronym{CDF}{
short = CDF,
long  = cumulative distribution function
}

\DeclareAcronym{IBDI}{
short = IBDI,
long  = inter-backscatter device interference
}

\DeclareAcronym{BPSK}{
short = BPSK,
long  = binary phase shift keying 
}

\DeclareAcronym{OFC}{
short = OFC,
long  = orthogonal frequency-code spread
}

\DeclareAcronym{TDMA}{
  short = TDMA,
  long = time division multiple access
}

\DeclareAcronym{CDMA}{
  short = CDMA,
  long = code division multiple access
}

\DeclareAcronym{IDFT}{
short = IDFT,
long  = inverse discrete Fourier transform
}

\DeclareAcronym{i.i.d}{
short = i.i.d,
long  = independent and identically distributed
}

\DeclareAcronym{NOMA}{
short = NOMA,
long  = non-orthogonal multiple access
}

\DeclareAcronym{RIS}{
  short = RIS ,
  long  = reconfigurable intelligent surface
}